\documentclass[referee]{raa}            

\usepackage{graphicx,times}             
\usepackage{natbib}
\usepackage{amssymb,amsmath}
\bibpunct{(}{)}{;}{a}{}{,}

\usepackage[pagebackref=true]{hyperref}

\begin{document}

  \title{Impact of Low-Earth Orbit Satellites on the China Space Station Telescope Observations
}

   \volnopage{Vol.0 (20xx) No.0, 000--000}      
   \setcounter{page}{1}          

   \author{Huai-Jin Tang 
      \inst{1,2}
   \and Xiao-Lei Meng
      \inst{1}
   \and Hu Zhan
      \inst{1,3}
   \and Xian-Min Meng
      \inst{1}
   \and You-Hua Xu
      \inst{1}
   }

   \institute{Key Laboratory of Space Astronomy and Technology, National Astronomical Observatories, Chinese Academy of Sciences, Beijing, 100101, China; {\it tanghj@nao.cas.cn}\\
        \and
             University of Chinese Academy of Sciences (UCAS), Beijing 100049, China\\
        \and
             Kavli Institute for Astronomy and Astrophysics, Peking University, Beijing, 100871, China\\
\vs\no
   {\small Received 20xx month day; accepted 20xx month day}}

\abstract{
It is projected that more than 100,000 communication satellites will be deployed in Low-Earth Orbit (LEO) over the next decade. These LEO satellites (LEOsats) will be captured frequently by the survey camera onboard the China Space Station Telescope (CSST), contaminating sources in the images. As such, it is necessary to assess the impact of LEOsats on CSST survey observations. We use the images taken by the Hubble Space Telescope (HST) in its F814W band to simulate $i$-band images for the CSST. The simulation results indicate that LEOsats at higher altitudes cause more contamination than those at lower altitudes. If 100,000 LEOsats are deployed at altitudes between 550 km and 1200 km with a 53-degree orbital inclination, the fraction of contaminated sources in a 150-s exposure image would remain below 0.50\%. For slitless spectroscopic images, the contaminated area is expected to be below 1.50\%. After removing the LEOsat trails, the residual photon noise contributes to relative photometric errors that exceed one-tenth of the total error budget in approximately 0.10\% of all sources. Our investigation shows that even though LEOsats are unavoidable in CSST observations, they only have a minor impact on samples extracted from the CSST survey. 
}
   \authorrunning{H.-J. Tang, X.-L. Meng, H. Zhan, X.-M. Meng \& Y.-H. Xu }            
   \titlerunning{Impact of LEO Satellites on CSST Observations}  
   \maketitle
%
%
\section{Introduction}           
\label{sect:intro}

The decreasing cost of deploying communication satellites, driven by advancements in rocket and spacecraft technology, is rapidly increasing the number of Low-Earth Orbit (LEO) satellites at altitudes ranging from 300 km to 1200 km. Several thousand LEO satellites (LEOsats) have already been in operation, with tens of thousands more anticipated to be launched over the next decade. Five major mega-constellation projects have been proposed: Starlink, OneWeb, Kuiper, Telesat, and GuoWang. Mega-constellations may further increase the number of LEOsats beyond $10^{5}$ within the next decade (\citealt{Venkatesan2020, Walker2020, Rawls2021, Bassa2022}). As a result, the impact of satellite constellations on optical astronomy is expected to intensify, leading to increased disruptions in astronomical observations and a degradation in data quality.

Mega-constellations have long been a subject of concern within the astronomical community. The concept of satellite constellations was initially introduced as 'Walker constellations' \citep{Walker1984}, and the launch of the Iridium satellites marked the beginning of their impact on astronomical observations \citep{James1998}. SpaceX deployed its first batch of 60 Starlink communication satellites in May 2019, intensifying concerns among astronomers regarding the potential impact of large satellite constellations on optical and near-infrared observations (\citealt{McDowell2020, Hainaut2020, Walker2020}). The Satellite Constellations 1 (SATCON1) conference systematically investigated the impact of the satellite constellations on optical astronomy and explored mitigation strategies to reduce the effects of bright LEOsats. Their primary focus was on assessing the impact on ground-based astronomy, particularly for wide-field surveys. SATCON1 recommended that satellites in the 550 km orbit should have an apparent magnitude fainter than $V_{\rm mag} = 7$ to minimize their visibility in astronomical observations \citep{Walker2020}. Beyond their effects on observational astronomy, the proliferation of LEOsats has also increased the hazard of space debris. To mitigate long-term risks, SATCON1 suggested that LEOsats should be deployed below 600 km to ensure deorbiting within 25 years. In contrast, satellites positioned at altitudes of 1200 km have deorbit times on the order of several centuries, contributing to the long-term accumulation of space debris in Earth’s orbital environment \citep{McDowell2020}.

The China Space Station Telescope \citep[CSST, ][]{Zhan2011, Gong2019, Zhan2021}, also known as the Xuntian Mission, is designed to carry out a large-scale multiband imaging and slitless spectroscopic survey. It is a 2-meter aperture telescope in LEO at an altitude of approximately 400 km, making it susceptible to interference from LEOsats. The CSST’s primary scientific objectives include studies of the Large-Scale Structure of the universe, strong and weak gravitational lensing, galaxy clusters, individual galaxies, and active galactic nuclei. Over its 10-year mission, the CSST will survey approximately 17,500 deg$^{2}$ of the sky, achieving high accuracy in both photometric redshifts (photo-z) and spectroscopic redshifts (spec-z) (\citealt{Cao2018, Gong2019, Zhou2021, Li2023, Miao2023}). Given its sensitivity and extensive sky coverage, it is crucial to assess the potential impact of LEOsats on CSST observations. A detailed study is therefore necessary to quantify and mitigate the effects of the LEOsat contamination on the telescope’s data quality.

The impact of LEOsats on space telescopes such as the CSST has not been studied as extensively as their effects on ground-based observatories. However, there are notable similarities in the ways LEOsats interfere with both types of telescopes. Bright satellite trails reduce the signal-to-noise ratio (SNR) and, in more severe cases, can lead to permanent data loss in contaminated regions. One key distinction is that LEOsats are severely out of focus for space telescopes, resulting in significantly wider trails than the size of the telescope’s Point Spread Function (PSF). Additionally, LEOsat trails exhibit broad, low surface brightness wings in images \citep{Walker2020}. If these trails are not properly masked, they can introduce systematic biases, particularly in weak gravitational lensing studies, potentially affecting cosmological analyses. Transient phenomena of LEOsats, such as flares and glints, pose additional challenges by contaminating observations of transient astrophysical events, including supernovae, gamma-ray bursts, and the electromagnetic counterpart of gravitational waves. The Zwicky Transient Facility (ZTF) identified 5,301 satellite streaks in the archival data collected from November 2019 to September 2021, attributing them to Starlink satellites \citep{Mroz2022}. The number of LEOsats visible above the horizon is greatest at latitudes near the orbital inclination, and the frequency of satellite crossings in an astronomical image is proportional to the total number of satellites, the exposure time, and the field of view (FoV) (\citealt{Tyson2020, McDowell2020, Hainaut2020, Kruk2023}). Moreover, when comparing two satellite constellations with an equal number of satellites, the expected number of high-altitude satellites passing through an image exceeds that of low-altitude satellites, due to their longer orbital paths and increased visibility duration (\citealt{Walker2020, Bassa2022, Kruk2023}).

There are key differences in the impact of LEOsats on ground-based and space telescopes. For space telescopes, factors such as seasonal variations, atmospheric scintillation, and atmospheric extinction are not relevant. In contrast, for ground-based telescopes like the Rubin Observatory, satellite constellations at 1200 km altitude can remain visible throughout the entire night during the summer \citep{Walker2020}. Additionally, atmospheric scintillation introduces irregularities in satellite trails, making it challenging to model their profiles accurately \citep{Bassa2022}. For space telescopes such as the CSST, close encounters with LEOsats raise safety concerns, necessitating preventive measures, even at the cost of observational time. A close encounter may also result in more intense sunlight reflection into the telescope. Moreover, the angular speed between LEOsats and a LEO space telescope is on average greater than that for ground-based telescopes, which affects the characteristics of satellite-induced contamination in images. When assessing the impact of LEOsats on a specific space telescope, it is necessary to consider various factors, including the telescope's optical system, orbital parameters, scientific objectives, observational schedule, and other mission-specific constraints. A comprehensive understanding of these factors is crucial for developing effective mitigation strategies to minimize the impact of LEOsat contamination on space telescope observations.

In this paper, we aim to evaluate the potential impact of satellite mega-constellations on CSST observations via simulations. The simulations contain two main components: (1) the characteristics of LEOsats in the CSST's FoV, and (2) the properties of LEOsat trails in CSST images. We then analyze their impact on sources that can be detected in the CSST survey. Finally, we explore potential mitigation strategies to minimize these effects. We extract from the simulations key parameters of LEOsats when they enter the CSST’s FoV, including their brightness, quantity, angular velocity, solar phase angle (Sun-target-observer), and other relevant factors. We then simulate the LEO satellite trails and incorporate their photon noise into the synthetic images. We use the Hubble Space Telescope (HST) images taken in its F814W band for the COSMOS project \citep{Leauthaud2007} to simulate the $i$-band image for the CSST. The analysis focuses on the affected regions, noise levels, changes in contaminated sources, and the relationship between distance and photometric error.

The structure of this paper is as follows. In Section 2, we describe the simulation parameters, including CSST’s observational conditions, the characteristics of LEOsats and their trails, and the method for generating simulated images. In Section 3, we present the simulation results, examining the properties of LEOsats crossing the CSST's FoV and their impact on contaminated sources, particularly in terms of increased measurement uncertainty. Finally, in Section 4, we summarize the significance and implications of our findings.

\section{Simulation}

The objective of this simulation is to model the characteristics of a Sun-illuminated LEOsat as they traverse the FoV of the CSST across different bands. The CSST is a forthcoming full-sky survey that is  currently in the planning stages. It is designed to conduct both photometric imaging and slitless spectroscopic surveys simultaneously (\citealt{Zhan2011, Gong2019, Zhan2021}). Unlike time-domain astronomy, the CSST’s observation schedule is optimized specifically to its scientific mission. This schedule directly influences the zenith angles of the telescope's pointing and determines the characteristics of Sun-illuminated LEOsats within the CSST’s FoV.

The parameters of LEOsats in our simulation relate to the ground-based simulation from the previous study \citep{Walker2020}. For consistency, we assume LEOsats with a size of 2 meters, as defined in earlier work \citep{Tyson2020}. We employ a simplified brightness model in which an observer on the ground at a distance of 550 km from the LEOsats would perceive their apparent magnitude in the V band as 7. We also incorporate an atmospheric extinction factor of 0.2 in the V band for consistency with ground-based simulations \citep{Patat2011}. To assess the impact of satellites at varying altitudes, we conduct simulations at orbital heights of 550 km and 1200 km. Following the SATCON1 study, we deploy $10^{4}$ satellites at each altitude, distributing 100 satellites across each of 100 orbital planes, with an orbital inclination set at 53 degrees \citep{Walker2020}. For the image simulation, we use two real observational catalogs — COSMOS and zCOSMOS — to replicate the CSST survey conditions. The magnitude limits of these catalogs are deeper than those projected for the CSST, ensuring a comprehensive analysis (\citealt{Zhan2011, Cao2018, Gong2019}). The details of this simulation are organized into three primary sections: characteristics of LEOsats, CSST observation conditions, and image simulation. Each of these components will be discussed thoroughly in the subsequent sections.

\subsection{Characteristics of LEOsats}

The defining characteristics of mega-constellations include their efficient utilization of limited orbital space, operation in low Earth orbit, use of high-inclination orbits for global coverage, and the management of high reflectivity to regulate satellite temperature. Starlink satellites feature a flat-panel design approximately 3 meters in length, with all components mounted on this surface. Their solar panels are positioned at a right angle to the flat panel \citep{McDowell2020}. They have also experimented with specialized darkening coatings to mitigate diffuse reflection. In contrast, OneWeb satellites utilize a pair of extended solar panels. For our simulation, we adopt a LEOsat size of 2 meters, consistent with parameters used in previous research \citep{Tyson2020}. The orbital configurations of five major proposed mega-constellations are summarized in Table 1, with the LEOsats parameters from our two altitude simulations appended at the end for reference (\citealt{Walker2020, Bassa2022}). 

\begin{table} 
\center
\caption{Orbit configurations} 
\begin{tabular}{lllll}
\hline
 &Mega-Constellation    &Altitudes (km)   &Inclinations (deg)   &Number of satellites\\
\hline
&SpaceX (Starlink)         &328 - 614    &30.0 - 148.0    &41926     \\
&OneWeb                        &1200     &40.0 - 87.9       &47844    \\
&Amazon (Kuiper)            &590 - 630    &33.0 - 51.9      &3236     \\
&Telesat                &1000-1325       &37.4 - 99.5     &1671     \\
&GuoWang (GW-A59, G60)       &500-1200          &30.0 - 85.0      &15000 \\
&Simulation LEOsats (550 km)      & 550     &53      &10000    \\
&Simulation LEOsats (1200 km)      & 1200    &53      &10000   \\
\hline
\end{tabular}
\end{table}

The apparent magnitude of LEOsats has been measured and normalized to a nominal orbital altitude of 550 km. A dataset comprising 281 observations of Starlink satellites yielded a mean V-band magnitude of 5.5 with a standard deviation of 1.0 \citep{Walker2020}. Another set of 353 observations resulted in an average GAIA G band magnitude of 5.5 $\pm$ 0.13, with a standard deviation of 1.12 \citep{Halferty2022}. The brightness of Qianfan LEOsats varies from a visual magnitude of 4 near the zenith to 8 when closer to the horizon \citep{Mallama2024}. The brightest known LEO communication satellite constellation, AST Space Mobile's BlueWalker 3, reaches a peak visual magnitude of 0.4 \citep{Nandakumar2023}. SpaceX has attempted various mitigation strategies to reduce satellite brightness. One such effort was Darksat, a Starlink satellite coated with a special darkening material. The observations indicated that Darksat reduced the Sloan $g'$ magnitude by 0.77 $\pm$ 0.05 mag \citep{Tregloan2020}. Additional reductions were observed in other bands: 50\% in Sloan $r'$, 42\% in Sloan $i'$, 32\% in NIR J, and 28\% in NIR Ks \citep{Tregloan2021}. However, due to thermal control issues, SpaceX discontinued Darksat and introduced VisorSats, which use deployable sun visors to mitigate reflectivity. The brightness of VisorSats was measured at a mean magnitude of 6.0, with a standard deviation of 0.79 \citep{Halferty2022}. Another study, based on 430 measurements, reported an average visual magnitude of 5.92 $\pm$ 0.04 \citep{Mallama2021}. The SATCON1 established a recommended brightness threshold for satellites, stating that they should have an apparent magnitude fainter than $V_{\rm mag} = 7.0$ to minimize their impact on astronomical observations \citep{Walker2020}. However, even if LEOsats dim to this level, their trails will still be approximately 100 times brighter than the sky background noise, introducing systematic errors that may constrain certain scientific objectives \citep{Tyson2020}. In our simulation, we adopt this threshold to construct an apparent magnitude model for LEOsats. In contemporary astronomy, multiband observations of LEOsats have led to substantial data accumulation and analysis (\citealt{Campbell2019, Hossein2023, Zhi2024}). By tracking satellite trajectories within observational images, researchers can identify satellite types more accurately (\citealt{Krantz2022, Kruk2023}). Furthermore, advancements in precise Bidirectional Reflectance Distribution Function (BRDF) models have improved satellite brightness modeling, enabling more refined assessments of their impact on astronomical observations (\citealt{Fankhauser2023, Lu2024}).

The accurate apparent magnitude model of LEOsats is dynamic and influenced by many parameters, including the solar phase angle (Sun-target-observer), satellite geometry, and orientation. Different satellite components exhibit varying levels of reflectivity that change over time. To reduce the number of simulation parameters, we adopt a simplified brightness model for LEOsats. As a first-order approximation, a satellite can be modeled as a uniform sphere with purely diffuse reflection, known as a Lambertian sphere (\citealt{Hainaut2020, Walker2020}). The solar phase angle is denoted by $\alpha$, and for a Lambertian sphere, the first-order approximation of the solar phase attenuation, $v_{\alpha}$, is given by $v_{\alpha}=(1+\cos(\alpha))/2$ \citep{Hainaut2020}. For an observer on the ground at a distance of 550 km from a LEOsat, the apparent V-band magnitude is assumed to be 7. For ground-based observations, the atmospheric extinction in the V band is assumed to be 0.2 magnitudes \citep{Patat2011}. When converting magnitudes from ground-based observations to those applicable for the CSST, this extinction value must be subtracted to account for the absence of atmospheric effects in space-based observations. The V band magnitude of the LEOsats (V$_{\rm sat}$) could be unified by a simple formula below \citep{Hainaut2020}, where $D_{\rm ist}$ represents the distance between the LEOsats and the CSST.

\begin{equation}
   {\rm V}_{\rm sat}= 7+ 5 {\rm log_{10}}(\frac{D_{\rm ist}}{550\mbox{ } \rm km}) - 2.5 {\rm log_{10}}(v_{\alpha}) - 0.2
   \label{eq:quadratic}
\end{equation}

The next step involves converting the V-band brightness of LEOsats, expressed in photon counts, to the corresponding CSST band brightness in the detector, measured in electron counts. To achieve this, we assume that LEOsats exhibit a solar-like spectral energy distribution, consistent with previous studies \citep{Tyson2020}. Subsequently, we account for the system efficiency, which encompasses multiple factors: the optical system efficiency, the intrinsic transmission curves of the filters, and the detector's quantum efficiency for each band (\citealt{Cao2018, Zhan2021, Cao2022}). By incorporating these elements, we derive the conversion coefficients that relate the photon count in the V band to the electron count in the seven photometric imaging bands (NUV, $u$, $g$, $r$, $i$, $z$, $y$) and the three slitless spectroscopic bands (GU, GI, GV). The calculated coefficients are as follows: 0.0349, 0.148, 0.859, 1.08, 1.06, 0.639, 0.184, 0.770, 2.62, 3.97, respectively. It is evident that the flux of LEOsat trails in the NUV band is significantly lower than in the other imaging bands. As a result, the NUV band is typically excluded from imaging analyses.


The following section details the simulation process for modeling the imaging cross-sectional profile and slitless spectrum profile of LEOsat trails at various distances. In our simulation, we do not account for the broad low surface brightness wings, as they are faint and challenging to model accurately. The first step involves generating the imaging off-focus PSFs for different bands at various distances. To achieve this, we use Zemax OpticStudio\footnote{\url{https://www.ansys.com/products/optics/ansys-zemax-opticstudio}} to simulate the CSST optical system. Table 2 presents the radius of 80\% energy concentration (REE80), which quantifies the imaging off-focus PSF sizes across different bands. Next, we simulate the imaging cross-sectional profile of LEOsat trails at varying distances. The TopHat model is used to represent the shape of LEOsats, considering their apparent image size within the FoV relative to distance. This model is then convolved with the corresponding imaging off-focus PSF at the same distance. By summing the resulting matrix along one direction, we obtain the imaging cross-sectional profile of the LEOsat trails. In Figure 1, we present the imaging cross-sectional profiles of LEOsat trails in the i band at various distances (left panel) and the Full Width at Half Maximum (FWHM) of LEOsat trails as a function of distance (right panel). For the slitless spectrum profile of LEOsat trails, we employ convolution to combine the cross-sectional profile  with the slitless spectrum profile. We assume that the zeroth-order spectrum contributes 10\% of the total flux, while the first-order spectrum accounts for 60\% of the total flux.

\begin{table} 
\center
\caption{The REE80 ($\rm \mu$m) of the off-focus PSF at various distances in different bands  }
\begin{tabular}{llllllll}
\hline
&Distance   &$u$ ($\rm \mu$m )  &$g$ ($\rm \mu$m) &$r$ ($\rm \mu$m) &$i$ ($\rm \mu$m) &$z$ ($\rm \mu$m) &$y$ ($\rm \mu$m)   \\
\hline
&500 km  &46.0  &46.8 &45.6 &46.3  &46.2 &46.4  \\
&1000 km &26.0  &26.1 &24.7 &27.2 &26.7 &26.7 \\
&2000 km  &15.8  &15.8 &14.9 &17.7  &19.3 &19.7  \\
\hline
\end{tabular}
\end{table}

\begin{figure}
	\includegraphics[width=0.48\columnwidth]{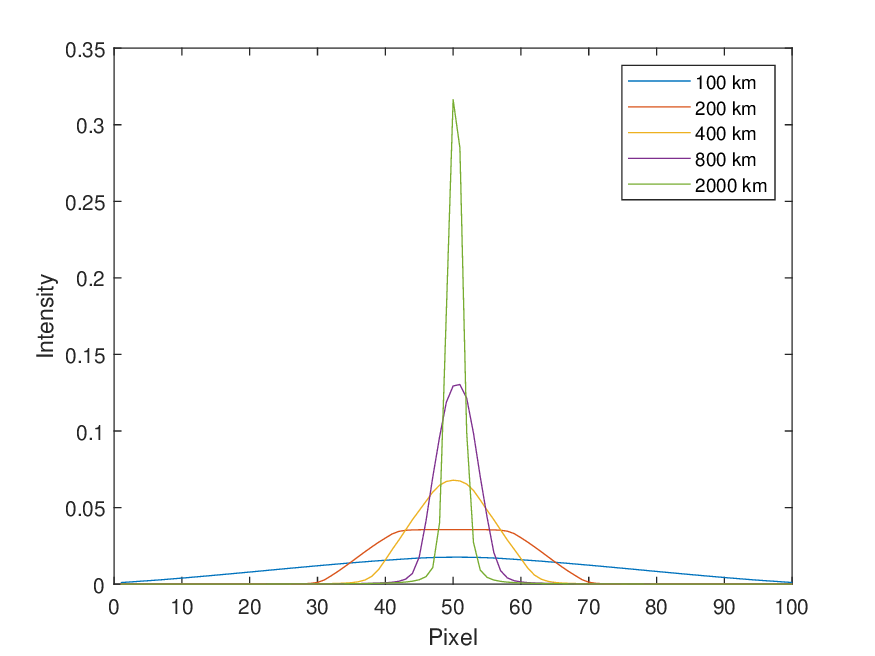}
	\includegraphics[width=0.48\columnwidth]{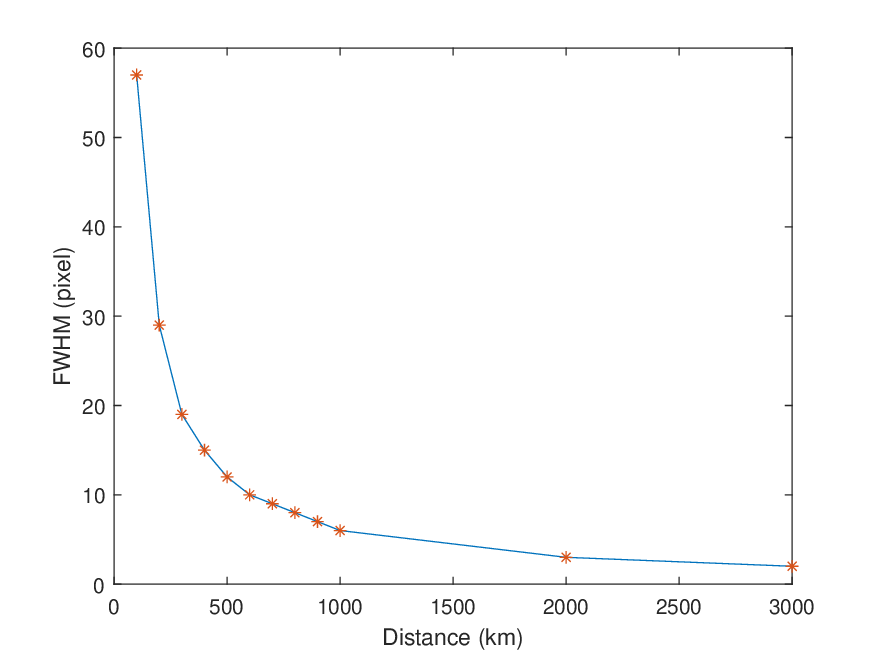}
 \caption{The left panel presents the cross-sectional profiles (in pixels) of off-focus LEOsat trails in the $i$-band at varying distances. The right panel illustrates the FWHM (in pixels) of the $i$-band LEOsat trails at different distances between the LEOsats and the CSST. The pixel size is 10 $\rm \mu$m. As the distance decreases, the trail profile becomes noticeably broader.
 }   
\label{fig:example_figure}
\end{figure}

\subsection{CSST Observation Conditions}

The simulated observation schedule of CSST is a critical factor influencing the observed characteristics of LEOsats in the CSST FoV. Several key considerations must be addressed when designing this schedule, with the primary factor being the telescope’s scientific objectives. The CSST orbits at an altitude of approximately 400 km in LEO, with an orbital inclination of 42.5 degrees and an orbital period of approximately 90 minutes. During its wide-field survey, the CSST will cover approximately 17,500 square degrees with an exposure time of 150 seconds per observation. Each band will revisit the same region of the sky two or four times over a ten-year period. Additionally, a deep survey covering an area of 400 square degrees is planned, with eight exposures per field, each with an integration time of 250 seconds (\citealt{Zhan2011,Gong2019}). The observation schedule is specifically designed to support its scientific mission, taking into account the positions of the Sun, Moon, Earth, and the Zodiacal light, among others. The observing direction is required to maintain an angular separation of at least 50 degrees from the Sun and 40 degrees from the Moon. Furthermore, the observed field must be at least 30 degrees away from the dark side of the Earth and at least 70 degrees from the illuminated side. Additional considerations include avoiding the South Atlantic Anomaly (SAA), ensuring optimal steering control and maintaining an appropriate angle between the Sun and the CSST's solar panels. These constraints are essential for achieving full coverage of the designated survey area. The distribution of zenith angles based on the simulated observation schedule exhibits two distinct peaks at approximately 30 degrees and 70 degrees, as evident from the count distribution. This pattern arises from two primary factors. First, the CSST operates differently on the dark side of the Earth compared to the illuminated side. On the illuminated side, selecting a smaller zenith angle helps mitigate the impact of scattered light from the Earth's surface. Second, in celestial coordinates, a higher zenith angle corresponds to a larger observable area, resulting in a lower frequency of observations at small zenith angles compared to larger zenith angles. Figure 2 presents the distribution of zenith angles based on the simulated observation schedule, which comprises 640,661 exposures of 150 seconds each over a ten-year period.

\begin{figure}
	\includegraphics[width=\columnwidth]{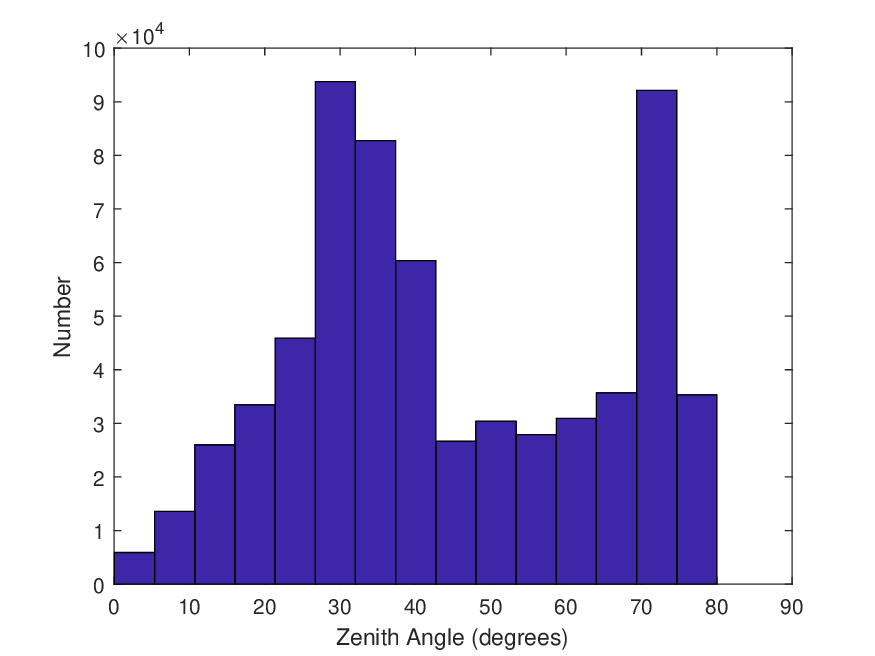}
 \caption{ The figure presents the distribution of zenith angles based on 640,661 simulated 150-second exposure observations. The simulated observation schedule exhibits two prominent peaks at approximately 30 degrees and 70 degrees. In the lower zenith angle regime, the number of observations increases with the zenith angle, reflecting the influence of observational constraints and survey strategy.
 }   
\label{fig:example_figure}
\end{figure}

The distribution of distances between LEOsats and CSST is directly influenced by the zenith angles of the simulated observation schedule. When the altitude of LEOsats remains constant, a larger zenith angle corresponds to an increased distance between the LEOsats and the CSST. Figure 3 displays the relationship between distance and zenith angle. Notably, the presence of two peak zenith angles in the observation schedule results in a bimodal distribution of distances.

\begin{figure}
\centering
	\includegraphics[width= \columnwidth]{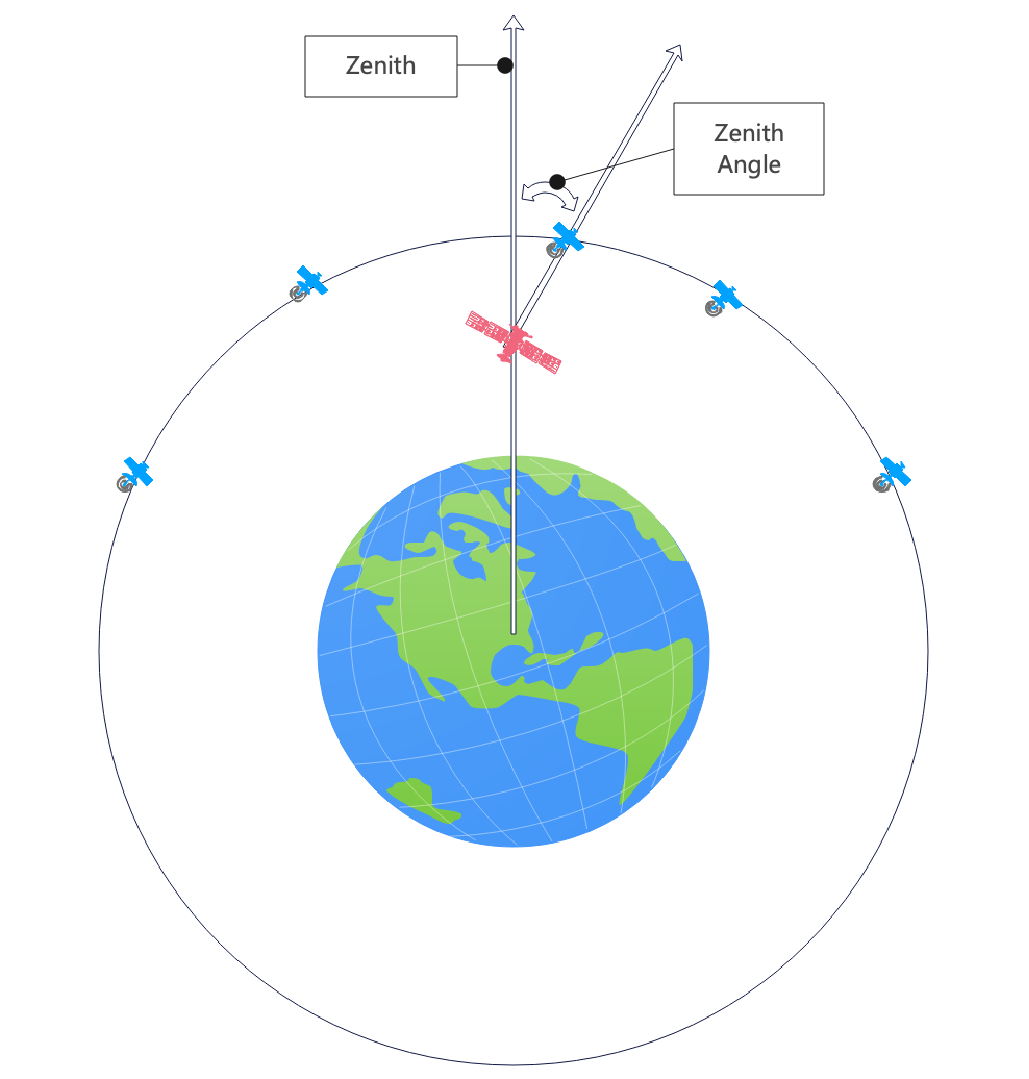}
\label{fig:side:a}
\caption{The figure illustrates the zenith angles of LEOsats as they pass through the CSST's FoV. It indicates that larger zenith angles correspond to greater distances between the LEOsats and the CSST.
}
\end{figure}

The CSST features a pixel size of 10 micrometers and a focal length of 28 meters, resulting in an angular resolution of approximately 0.074 arcseconds per pixel. Its FoV is approximately 1.1 square degrees. The CSST camera consists of 30 CCDs, each containing over $9000 \times 9000$ pixels, and supports observations across seven photometric imaging filters (NUV, $u$, $g$, $r$, $i$, $z$, $y$) as well as three slitless spectroscopic filters (GU, GI, GV). The background noise levels ($B_{\rm sky}$) for the photometric imaging filters are as follows: 0.003, 0.017, 0.16, 0.20, 0.21, 0.13, and 0.037 e$^{-}$s$^{-1}$pixel$^{-1}$ for NUV, $u$, $g$, $r$, $i$, $z$, and $y$ bands, respectively \citep{Cao2018, Zhan2021,Cao2022}. The measured dark current ($N_{\rm D}$) is approximately 0.02 e$^{-}$s$^{-1}$pixel$^{-1}$, while the read noise ($N_{\rm R}$) is approximately 5 e$^{-}$pixel$^{-1}$. The root mean square (RMS) noise ($\sigma_{\rm CSST}$) of a single pixel, in the absence of sources, is given by the formula below, where $N_{\rm sky}$ represents the background noise per pixel and $t$ is the exposure time. Using the $i$-band background noise as an example, the RMS noise per pixel is 7.5 e$^{-}$ for a 150-second exposure and 8.7 e$^{-}$ for a 250-second exposure.

\begin{equation}
\sigma_{\rm CSST}=\sqrt{(N_{\rm sky}+N_{\rm D})*t+N_{\rm R}^{2}}
\end{equation}

\subsection{Image Simulation} 

In the image simulation process, the first step is to generate simulated images based on the observation conditions of the CSST. To simulate the $i$-band image for the CSST, we utilize the HST F814W COSMOS image database as the basis for the simulation. The simulation procedure comprises three key steps:

\begin{itemize}
    \item Resampling the Image: Adjusting the image sampling frequency from the COSMOS database to match the resolution and pixel scale of the CSST.
    \item Calibrating the Photometric System: Adjusting the collected flux to account for differences in the collecting area and system efficiency between COSMOS and CSST.
    \item Incorporating Background Noise: Adjusting the background noise to reflect the conditions of a single simulated CSST exposure.
\end{itemize}

In the process of adjusting the image sampling frequency from COSMOS to CSST, it is essential to account for differences in spatial angular resolution. The CSST has a pixel scale of 0.074 arcseconds per pixel, which is approximately 2.5 times that of COSMOS (0.03 arcseconds) \citep{Leauthaud2007}. Given that the PSF image sizes in both systems are roughly comparable \citep{Zhan2021}, the resampling process involves two key steps: first, each COSMOS pixel is subdivided into a $2 \times 2$ grid of smaller pixels, and subsequently, a region of $5 \times 5$ pixels is combined into a single CSST pixel. This approach ensures that the sampling of the simulated image aligns with the optical and detector characteristics of the CSST, preserving spatial resolution while accurately representing the observational conditions.

In the process of adjusting the flux of sources in the image, this step ensures that the collecting area and system efficiency are properly scaled from COSMOS to CSST. It requires the application of conversion factors to transform a COSMOS F814W band image into a corresponding CSST band image with the same exposure time. The scaling conversion factor, {\it F}, is defined by the formula below, where $A_{\rm CSST}$ and $A_{\rm HST}$ represent the mirror areas responsible for photon collection, $T_{\lambda,\rm CSST}$ denotes the system throughput curve of a given CSST observing band, $T_{\lambda,\rm F814W}$ corresponds to the HST F814W band system throughput curve. Additionally, $f_\lambda$ represents the average SED of celestial objects. For simplicity, we assumed a constant average SED to approximate the combined contributions of various celestial objects.

\begin{equation}
F= \frac { A_{\rm CSST} \cdot \int{f_\lambda \cdot T_{\lambda,\rm CSST} \lambda } \mathrm{d} \lambda} { A_{\rm HST} \cdot \int{f_\lambda \cdot T_{\lambda,\rm F814W} \lambda}  \mathrm{d} \lambda},
\end{equation}

In image simulation, it is essential to incorporate noise to accurately replicate the observing conditions of the CSST. The magnitude limit of the CSST simulated image is not deeper than that of COSMOS, as the COSMOS dataset consists of coadded images from multiple exposures. To appropriately introduce noise into the CSST simulated images, such as in the i band, we apply the formula shown below. Here, $\sigma_{{\rm add},i}$ represents the additional noise required to simulate the CSST image, while $\sigma_{{\rm scaled},i}$ denotes the background noise of the COSMOS-scaled image, which has already been adjusted by the scaling conversion factor.

\begin{equation}
    \sigma_{{\rm add},i} = \sqrt{\sigma^2_{{\rm CSST},i} - \sigma^2_{{\rm scaled},i}},
\end{equation}

The source density in the CSST simulated image is lower than that in the COSMOS image. The magnitude limit of COSMOS in the F814W filter is 26.5, with a galaxy density of approximately 66 galaxies per arcmin$^{2}$ \citep{Leauthaud2007}. In our simulated images with a 150-second exposure time, the source density in the $i$-band, including both stars and galaxies, is approximately 32 sources per arcmin$^{2}$. After selecting sources with a SNR greater than 5, the source density is approximately 25 sources per arcmin$^{2}$. The parameters used in the simulated image are summarized in Table 3. The first row of the table provides the sky background noise ($B_{\rm sky}$) for each band, measured in electron count per pixel per second (\citealt{Ubeda2011, Cao2018}). The second row presents the LEOsat conversion factors for translating the V-band photon flux of LEOsats into the electron counts per second in the CCD for each band. The third row represents the source conversion factors used to transform a COSMOS F814W band image into a CSST image at an equivalent exposure time. The fourth row lists the magnitude limits for each band, corresponding to the CSST wide survey with 150-second exposures \citep{Liu2024}.

\begin{table}
\center
\caption{The parameter used in the simulation to convert images from COSMOS to CSST for each band } 
\begin{tabular}{llllllllll}
\hline
 &Band    &NUV    &$u$  &$g$   &$r$  &$i$  &$z$  &$y$  \\
\hline

&Sky Background noise (e$^{-}$pixel$^{-1}$s$^{-1}$)  &0.003 &0.017 &0.16 &0.20 &0.21 &0.13 &0.037   \\
&LEOsat conversion factors$^{1}$ (V band)   &0.0349  &0.148 &0.859 &1.08 &1.06 &0.639 &0.184 \\
&Source conversion factors$^{2}$ (F814W) &0.235 &0.344  &0.878 &0.735 &0.615 &0.360  &0.104 \\
&Magnitude limit (150 s)  &24.53 &24.90 &25.76 &25.48 &25.32 &24.77 &23.56 \\
\hline
\end{tabular}
\caption*{\small{The LEOsat conversion factors$^{1}$ (V band) are used to transform V-band photon counts from ground-based observatories and SATCON1 into CSST band electron counts. Similarly, the source conversion factors$^{2}$ (F814W) convert electron counts from the COSMOS HST F814W band to CSST band electron counts.}}
\end{table}

The next step involves incorporating LEOsat trails into the simulated images and analyzing the contaminated sources. We utilize the Python library SEP to detect and quantify variations in contaminated sources \citep{Barbary2016}. In the source extraction procedure, we apply a $3 \times 3$ pixel filtering kernel and set the threshold to one times the RMS of the un-convolved image background noise. The DETECT\_MINAREA threshold is set to 13 pixels, while the window function size for source extraction is defined as 16 pixels. Contaminated sources are defined as those exhibiting magnitude errors caused by LEOsat trails when photometry is measured within a 2.5 Kron radius, with a magnitude error threshold of 0.001. Figure 4 displays a segment of a LEOsat trail within the simulated image, covering $2000 \times 2000$ pixels. The pixel scale and noise level have been calibrated to match the CSST's parameters in the $i$-band.

\begin{figure}
\centering
	\includegraphics[width=\columnwidth]{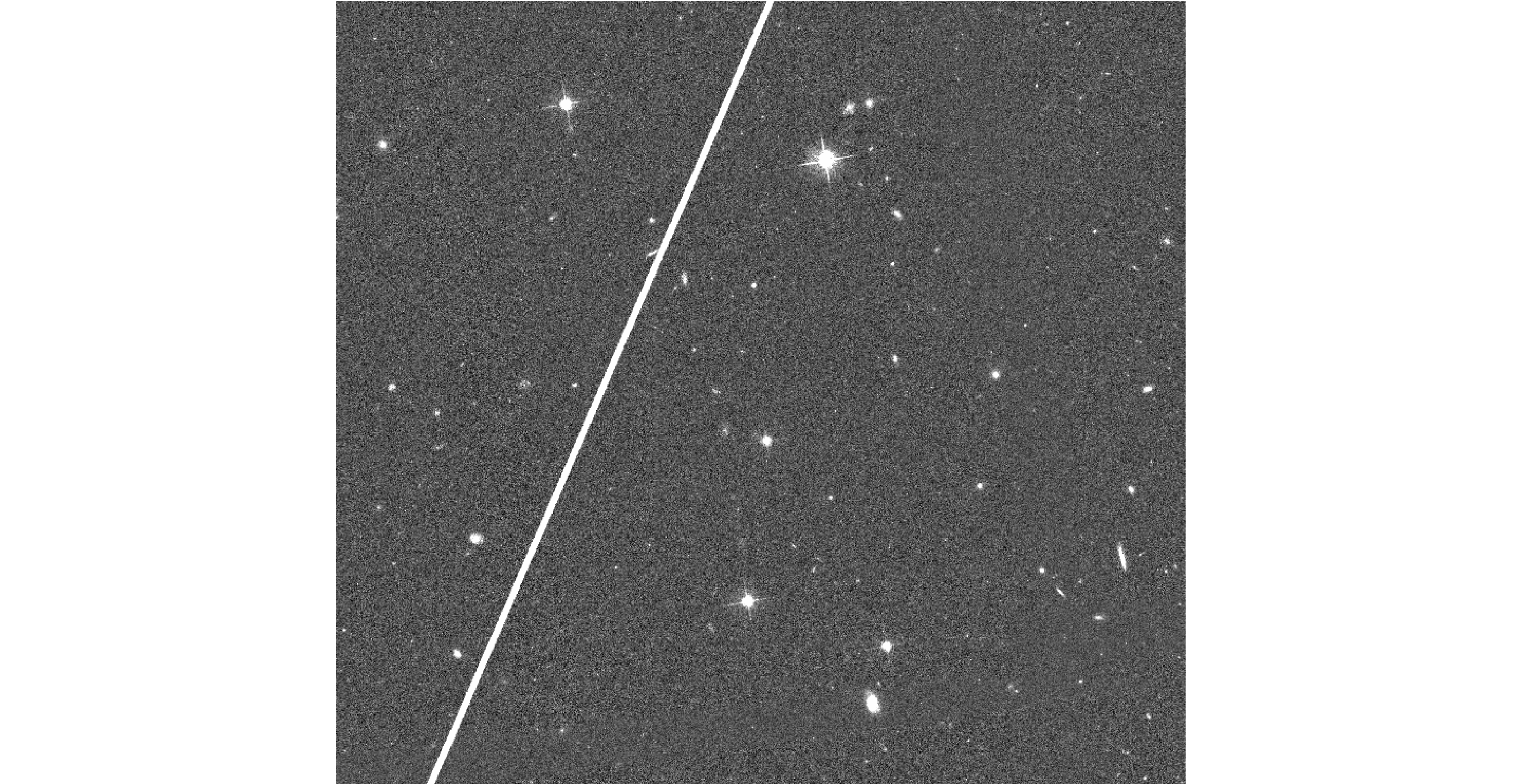}
\label{fig:side:a}
\caption{A segment of a LEOsat trail within the simulated image (length = 2000 pixels). This section contains approximately 200 sources, including both stars and galaxies. The simulation is based on COSMOS data, with the pixel scale and noise level calibrated to match the observational parameters of the CSST.}
\end{figure}

We assessed the impact of LEOsat trails by quantifying the number of contaminated sources and the errors introduced in both stars and galaxies. The parameters of the LEO satellite trails present in the simulated CSST images were obtained, including their quantity, brightness, length, and profile. To characterize the shape of sources, we adopt the second brightness moments ($Q_{ij}$) of unweighted sources, defined as: $Q_{ij} = (\int d^2 x \, I(x) \, x_i x_j)/(\int d^2 x \, I(x))$, where \( x_1 \) and \( x_2 \) correspond to the \( x \) and \( y \)-coordinates, respectively, \( I(x) \) represents the light profile of the source, and \( x = 0 \) denotes the center of the source. The ellipticity, used to define the shape, is given by the following formula, where $e_{1}$ represents the ellipticity along the principal axis, and $e_{2}$ represents the ellipticity along the diagonal.
\begin{equation}
   e_{1}= \frac{Q_{11}-Q_{22}}{Q_{11}+Q_{22}}
   \label{eq:quadratic}
\end{equation}

\begin{equation}
   e_{2}= \frac{2 \cdot Q_{12}}{Q_{11}+Q_{22}}
   \label{eq:quadratic}
\end{equation}


In the simulation of slitless spectroscopy, we extract source parameters from the zCOSMOS and Zurich databases, including brightness, size, and shape. The dataset contains approximately 20,000 galaxies, with a source density of about 3.3 sources per arcmin$^{2}$ \citep{Lilly2007}. We employ GalSim\footnote{\url{https://github.com/GalSim-developers/GalSim}} to generate simulated slitless spectroscopic images. Figure 5 illustrates a LEOsat trail within one of CSST-simulated slitless spectroscopic images. The zCOSMOS survey, with an I-band magnitude limit of I$_{\rm AB} < 22.5$, covers 1.7 deg$^{2}$ within the COSMOS ACS field.  We select the central region of the zCOSMOS survey to generate simulated images, resulting in an average source density of approximately 4.2 sources per arcmin$^{2}$. For example, in the GI band, the distance from the zeroth-order spectrum to the first-order spectrum is approximately 5440 $\rm \mu$m, while the length of the first-order spectrum measures around 3320 $\rm \mu$m. Each slitless spectroscopic image is generated using a pair of opposing slitless gratings, leading to the formation of five distinct bright bands within the spectroscopic image.

\begin{figure}
\centering
	\includegraphics[width=\columnwidth]{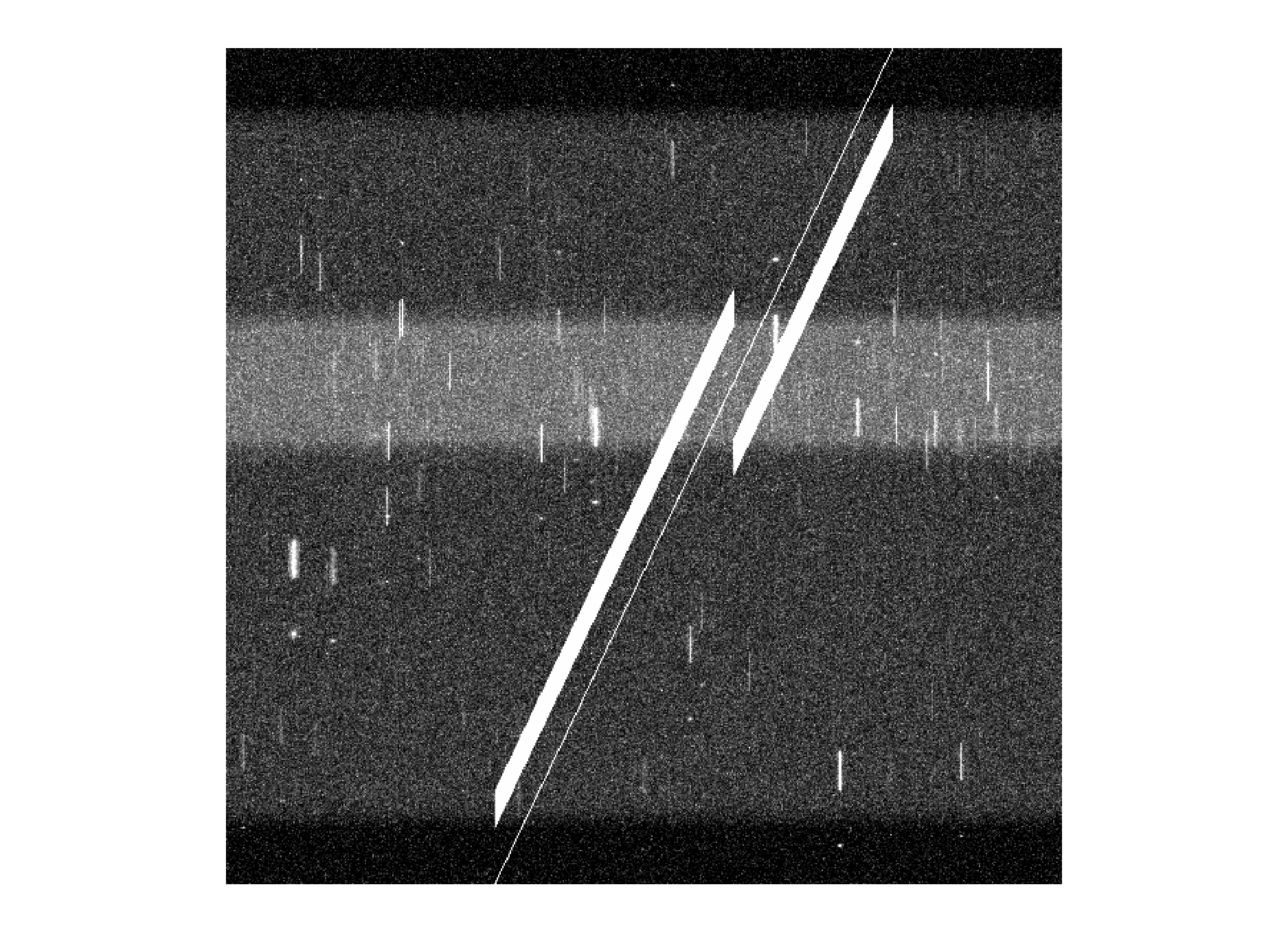}
\label{fig:side:a}
\caption{A LEOsat trail within the CSST-simulated slitless spectroscopic image spans a length of 9000 pixels. Each slitless spectroscopic image is generated using a pair of opposing slitless gratings. The dark regions at the upper and lower edges correspond to the areas between the zeroth-order and first-order slitless spectra of the background. Background photons are dispersed along the direction of spectral dispersion, resulting in greater brightness in the central region compared to the edges. At the center, the overlap of the two slitless spectroscopic images produces a significantly brighter band. }
\end{figure}

\section{Result and Discussion}

In the simulation, the observation schedule includes 703,657 exposures, comprising 640,661 exposures of 150 seconds, 54,608 exposures of 250 seconds, and a smaller number of exposures of varying durations. The analysis considers all LEOsats, regardless of whether they are illuminated by sunlight. The percentage distribution of specific LEOsats numbers passing through CSST's FoV during each exposure for a sample of $10^{4}$ LEOsats at different altitudes is presented in Table 4. The results demonstrate a positive correlation between the expected number of LEOsats crossing the CSST's FoV per exposure and both the exposure time and the satellites' altitude during observations. The characteristics of LEOsats as they pass through the CSST's FoV at altitudes of 550 km and 1200 km are summarized in Table 5 and Figure 6. These statistical results are derived from 640,661 simulated 150-second exposures. The results indicate that LEOsats at higher altitudes exhibit a greater proportion of illumination and significantly lower angular speeds compared to their low-altitude counterparts. Additionally, the bright trails produced by high-altitude satellites appear fainter than those generated by low-altitude satellites.

\begin{table} 
\center
\caption{
The percentage distribution of LEOsat numbers passing through the CSST's FoV} 
\begin{tabular}{lllllllll}
\hline
 &LEOsats Altitude  &Exposure Time    &0 LEOsats  &1 LEOsats  &2 LEOsats   &3 LEOsats  &  Expected Value   \\
\hline
&550 km  &150 s    &83.3$\%$  &14.8$\%$    &1.53$\%$  &0.25$\%$  &0.193    \\
&1200 km &150 s   &60.3 $\%$  &30.6$\%$   &6.99$\%$ &1.48$\%$    &0.515   \\
\hline
&550 km    &250 s    &77.9$\%$   &18.7$\%$   &2.84$\%$  &0.39$\%$  &0.266   \\
&1200 km   &250 s   &47.1$\%$   &36.5$\%$   &11.6$\%$  &2.96$\%$  &0.769     \\
\hline
\end{tabular}
\caption*{\small{The third to sixth columns display the proportions of LEOsats passing through the FoV during each exposure. The seventh column presents the expected number of LEOsats crossing the FoV per exposure. }}
\end{table}


\begin{table} 
\center
\caption{Characteristics of two altitude LEOsats in the CSST's FoV  }
\begin{tabular}{llll}
\hline
&LEOsats Altitude      &550 km   &1200 km  \\
\hline
&Illuminated LEOsats Ratio  & 66.0$\%$     &73.9$\%$   \\
&Length of One Trail (deg)  &0.93           &0.90    \\ 
&Distance (km)              &492.4        &1588.1    \\   
&Solar Phase Angle (deg)    &97.72 $\pm$ 30.74      &98.38 $\pm$ 28.58 \\
&$i$-Band Magnitude           &7.39 $\pm$ 1.40   &10.01 $\pm$ 1.10 \\
&Angular Speed (deg/s)      &1.26          &0.35  \\ 
&$i$-Band Flux (electron/pixel-length)  &1607.0   &467.07  \\ 
\hline
\end{tabular}
\caption*{\small{The illuminated LEOsats ratio  refers to the proportion of LEOsats which are illuminated by sunlight among all those passing through the CSST's FoV during 150-second exposures. Length of one trail represents the average length of each bright trail within the CSST's FoV. Distance denotes the average distance from illuminated LEOsats to the CSST detector. Solar phase angle provides the average and standard deviation of the solar phase angle for illuminated LEOsats. The $i$-band magnitude refers to the average and standard deviation of the $i$-band apparent magnitude for illuminated LEOsats. Angular speed represents the average angular speed of illuminated LEOsats within the CSST's FoV. The $i$-band flux indicates the average $i$-band flux of LEO satellite trails, expressed as the electron count per pixel length along the satellite trail on the CCD.}}
\label{Table:NBstart}
\end{table}

\begin{figure}
\centering
\includegraphics[width=0.48\linewidth]{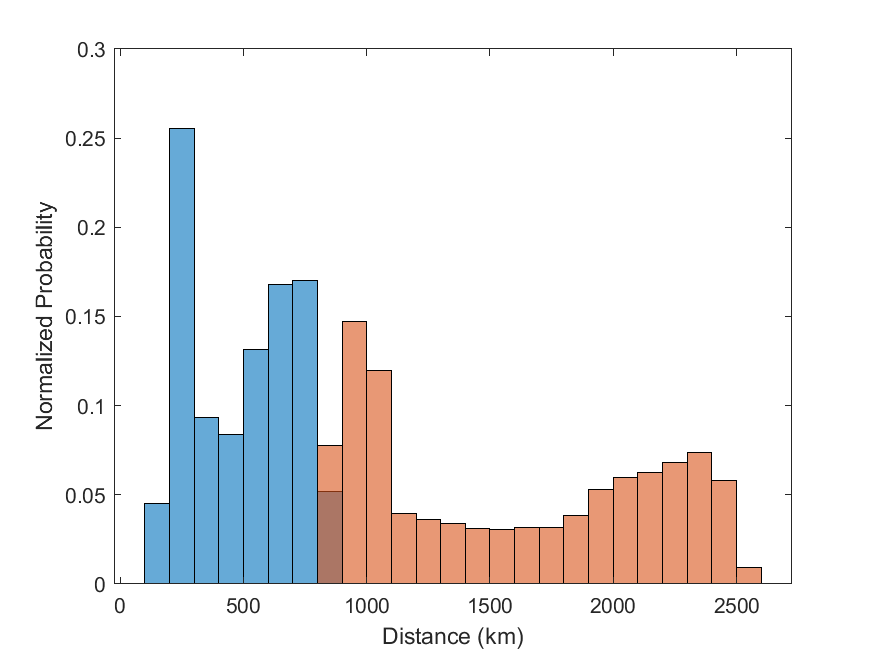}
\label{fig:side:a}
\hspace{0.01\linewidth}
\includegraphics[width=0.48\linewidth]{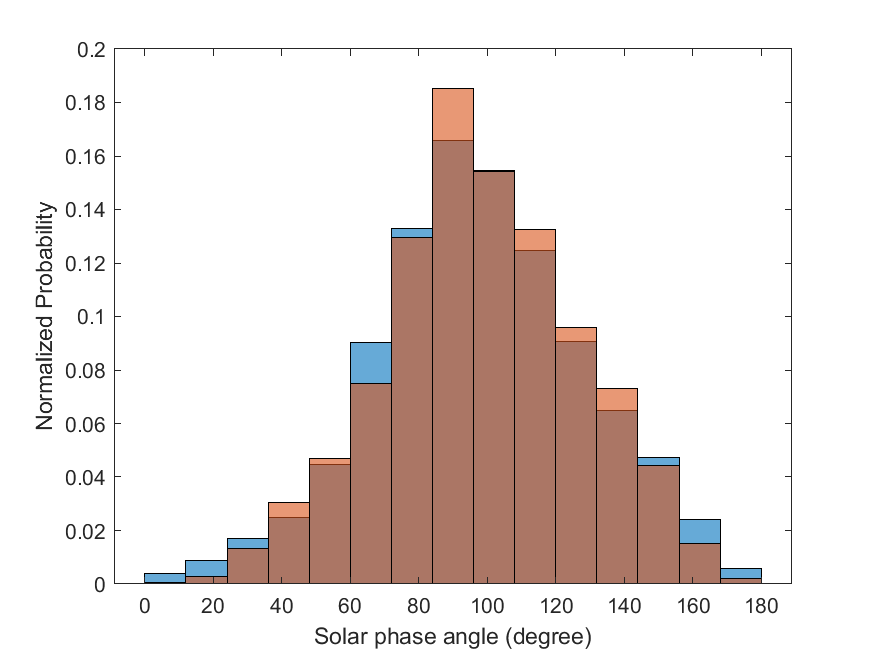}
\label{fig:side:b}
\includegraphics[width=0.48\linewidth]{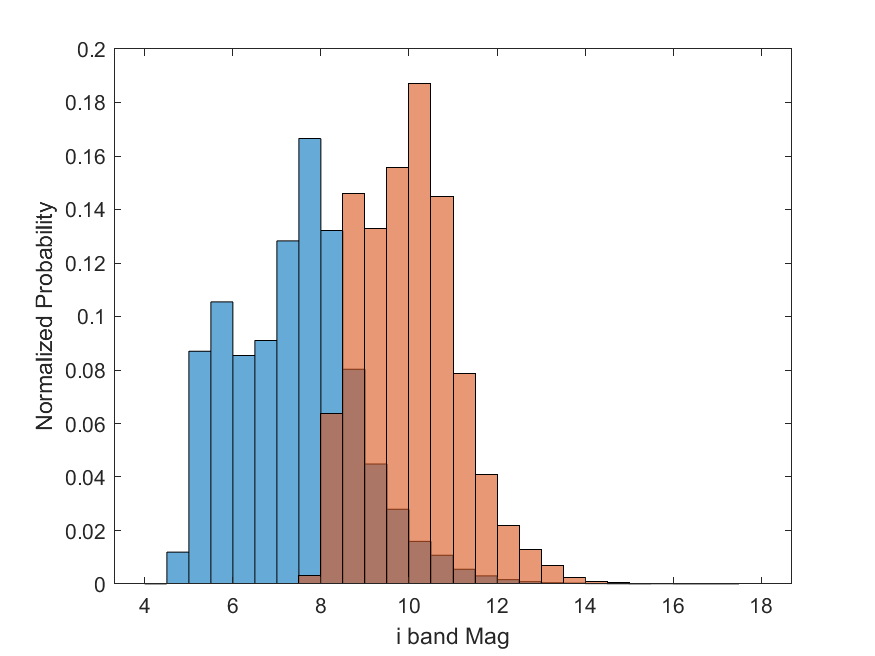}
\label{fig:side:c}
\hspace{0.01\linewidth}
\includegraphics[width=0.48\linewidth]{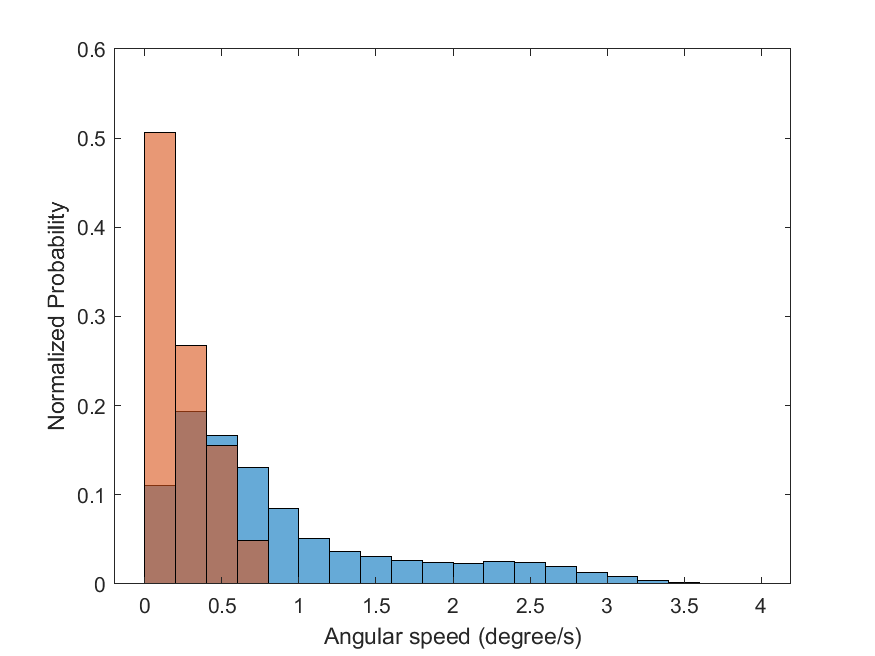}
\label{fig:side:d}
\includegraphics[width=0.48\linewidth]{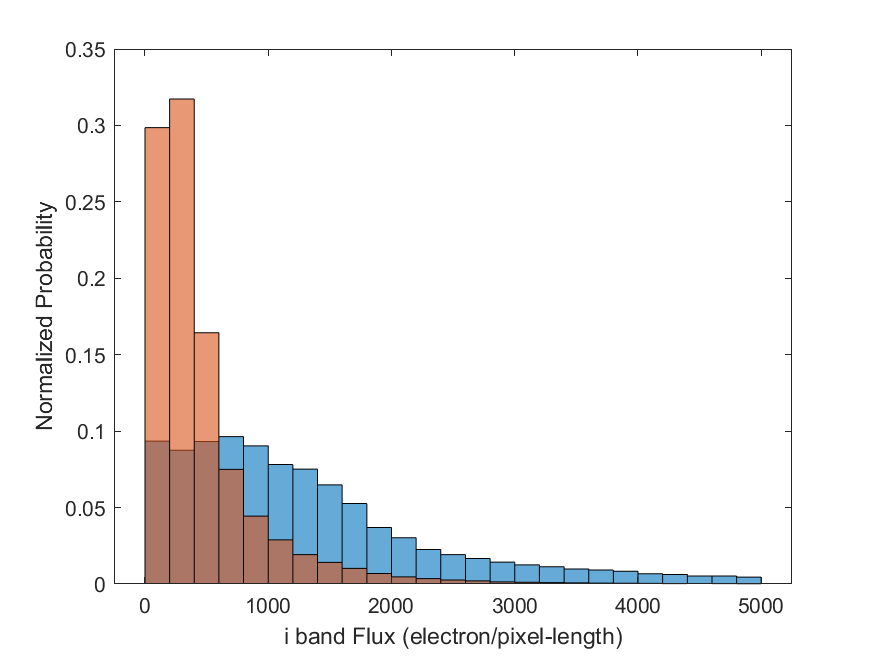}
\label{fig:side:e}
\caption{These figures illustrate the characteristics of illuminated LEOsats at altitudes of 550 km (blue) and 1200 km (brown) as they cross the CSST's FoV. The descriptions for each panel are as follows: the upper-left panel shows the distance of LEOsats from the CSST. The upper-right panel displays the solar phase angle of LEOsats; the lower-left panel presents the $i$-band magnitude of LEOsats; and the lower-right panel illustrates the angular speed of LEOsats within the CSST's FoV. The final panel is the $i$-band electron count per pixel length along the satellite trail on the CCD.
}
\end{figure}

The impact of LEOsat trails on CSST's scientific objectives can be classified into three main aspects: data loss, satellite streaks, and error introduction. Data loss occurs due to the reduction of effective observation time, field contamination from excessive bright trails, and crosstalk effects caused by overly bright trails or flares. Crosstalk effects caused by excessively bright LEO satellite trails cannot be effectively corrected. However, when the peak flux of a satellite trail is below approximately 5000–10,000 e$^{-}$, a preliminary nonlinear correction algorithm can successfully mitigate most variations in crosstalk coefficients \citep{Tyson2020}. Given that the majority of LEO satellite trails in CSST observations exhibit peak fluxes below 1000 e$^{-}$, the impact of crosstalk on CSST data is expected to be less significant compared to other space telescopes \citep{Walker2020}. Flares originate from specular reflections off specially designed spacecraft surfaces. In extreme cases, these flares can reach a magnitude of zero, rendering the entire exposure unusable due to uncorrectable crosstalk effects or detector saturation \citep{Tyson2020}. SATCON1 recommends improving the precision and completeness of laboratory-measured Bidirectional Reflectance Distribution Function (BRDF) models to better characterize these reflections. Additionally, maintaining an up-to-date and highly accurate satellite ephemeris is crucial for predicting and mitigating the impact of such flares on astronomical observations \citep{Walker2020}.

Satellite streaks appear as spurious signals triggered by LEOsats passing through the telescope’s FoV. The availability of precise satellite ephemeris can help mitigate these disruptions. The expected number of LEOsats appearing in a single CCD image is relatively low. For a population of $10^{5}$ LEOsats, 150 seconds exposure time, the expected values per CCD are approximately 0.21 for bright satellites at an altitude of 550 km and approximately 0.64 for those at 1200 km. LEOsats in orbits near the CSST traverse the FoV rapidly during close encounters, and their brightness and sharpness can introduce false signals. Fake events in astronomical images can be identified. In the ZTF data processing system, they utilized the CREATETRACKIMAGE software to detect satellite streaks (\citealt{Laher2014,Mroz2022}). Additionally, CelesTrak offers 'supplemental Two-Line Elements (TLEs)' derived from SpaceX's latest Starlink ephemeris data, enabling satellite position prediction with a precision of 500 meters \citep{Mroz2022}. Finally, they can verify whether the satellite passes through the telescope's FoV during the exposure.

Error introduction is a critical concern in this study, as contamination from satellite trails can introduce systematic uncertainties into astronomical observations. We aim to quantify the increase in errors within image simulations and assess their impact on the accuracy of scientific measurements in the following sections.

\subsection{Result of Image Simulation: error introduction}

To evaluate the impact of LEOsat trails on astronomical sources in simulated images, we conducted a detailed analysis of 2000 simulated trails. Although most LEOsat trails can be identified and removed, the residual photon noise from these trails remains in the images and may influence scientific measurements. This residual noise can introduce systematic uncertainties in key parameters such as photometry, centroid determination, and morphological shape measurements. The results of this analysis are summarized in Tables 6 through 8, which present the changes in the distribution of affected sources due to either the presence of LEOsat trails or the photon noise left after trail removal ('repair'). These tables show the level of contamination in terms of photometric accuracy, centroid displacement, and shape distortion.

\subsubsection{Magnitude, centroid and shape error introduction} 

\begin{table}
\center
\caption{The percentage of contaminated sources exhibiting magnitude error increases beyond specified thresholds
}
\begin{tabular}{llllll}
\hline
&Magnitude error &550 km     &1200 km    &550 km (repair)    &1200 km (repair)   \\
\hline
&$\ >$0.1   &0.065$\%$ &0.12$\%$ &0.010$\%$ &0.011$\%$\\
&$\ >$0.05  &0.072$\%$ &0.14$\%$ &0.021$\%$ &0.029$\%$\\
&$\ >$0.01  &0.10$\%$ &0.22$\%$ &0.051$\%$ &0.089$\%$\\
&$\ >$0.005 &0.12$\%$ &0.28$\%$ &0.069$\%$ &0.13$\%$\\
&$\ >$0.001 &0.18$\%$ &0.45$\%$ &0.14$\%$ &0.33$\%$\\
\hline
\end{tabular}
\caption*{\small{The percentage of contaminated sources with magnitude error increases exceeding the threshold due to contamination from $10^{5}$ LEOsats at orbital altitudes of 550 km or 1200 km, in a $9000 \times 9000$ pixel $i$-band image obtained during a 150-second exposure. These errors arise either from the direct influence of bright LEOsat trails or from residual photon noise remaining after the trails have been removed ('repaired'). The first column of the table lists the threshold values for acceptable measurement errors. The second and third columns indicate the fraction of contaminated sources with error values exceeding these thresholds, expressed as a proportion of the total number of sources in a $9000 \times 9000$ pixel image. Post-repair results demonstrate a noticeable reduction in contamination, although some residual effects from photon noise remain.}}
\label{Table:NBstart}
\end{table}

The results indicate that the fraction of contaminated sources within a single image remains relatively small. Based on the analysis of 2000 simulated satellite trails, we find that for LEOsats at an altitude of 550 km with a 150-second exposure, a satellite trail spanning one degree in angular length contaminates, on average, approximately 144 sources. For satellites at an altitude of 1200 km under the same exposure conditions, the corresponding number of contaminated sources is approximately 131. The image size (corresponding to a length of 9000 pixels) covers approximately 0.034 square degrees, while the CSST FoV spans 1.1 square degrees. An $i$-band image with a 150-second exposure contains approximately 3900 sources with SNR greater than 3. Approximately 3000 sources with a SNR greater than 5 are selected as the sample for analysis.

Assuming a population of $10^{4}$ LEOsats at an altitude of 550 km, the expected number of illuminated satellites crossing the CSST FoV during a 150-second exposure is about 0.127. The average length of a bright trail across the FoV is approximately 0.12 degrees, translating to an expected trail length of 0.0037 degrees per CCD. Under these conditions, the fraction of contaminated sources in a single CCD image is approximately 0.018\%.

For satellites at 1200 km, the expected number of illuminated satellites crossing the FoV during a 150-second exposure increases to approximately 0.381. The total bright trail length across the FoV is about 0.34 degrees, yielding an expected trail length of 0.010 degrees per CCD. In this case, the corresponding contamination fraction is approximately 0.045\%.

These findings suggest that LEOsats at higher altitudes result in a greater degree of source contamination. However, even with $10^{5}$ satellites at orbital altitudes of either 550 km or 1200 km, the fraction of contaminated sources within a single image remains below 0.50\%. Similar trends are observed for errors in photometry, centroid estimation, and shape measurements.


\begin{table} 
\center
\caption{The percentage of contaminated sources exhibiting centroid error increases beyond specified thresholds}
\begin{tabular}{llllll}
\hline
&Centroid Error (pixel)     &550 km     &1200 km    &550 km (repair)    &1200 km (repair) \\
\hline
&$\ >$ 0.1   &0.077$\%$ &0.15$\%$  &0.050$\%$   &0.091$\%$    \\
&$\ >$ 0.05  &0.090$\%$ &0.19$\%$  &0.067$\%$   &0.13$\%$    \\
&$\ >$ 0.01  &0.11$\%$ &0.25$\%$  &0.098$\%$   &0.22$\%$    \\
&$\ >$ 0.005 &0.13$\%$ &0.29$\%$  &0.11$\%$   &0.26$\%$    \\
& $\ >$0.001 &0.18$\%$ &0.47$\%$  &0.17$\%$   &0.46$\%$     \\
\hline
\end{tabular}
\caption*{\small{Same as Table 6, for the centroid error.}}
\label{Table:NBstart}
\end{table}

\begin{table} 
\center
\caption{The percentage of contaminated sources exhibiting ellipticity error ('repair') increases beyond specified thresholds
}
\begin{tabular}{llllll}
\hline
&Ellipticity Error &($e_{1}$, $e_{2}$ 550 km ) &($e_{1}$, $e_{2}$ 550 km repair)  &($e_{1}$, $e_{2}$ 1200 km)  &($e_{1}$, $e_{2}$ 1200 km repair)   \\
\hline
& $\ >$ 0.1   &(0.052$\%$, 0.069$\%$) &(0.10$\%$, 0.14$\%$)  &(0.028$\%$, 0.046$\%$)   &(0.048$\%$, 0.086$\%$)\\
& $\ >$ 0.05  &(0.068$\%$, 0.085$\%$) &(0.14$\%$, 0.18$\%$)  &(0.044$\%$, 0.065$\%$)   &(0.083$\%$, 0.13$\%$) \\
& $\ >$ 0.01  &(0.10$\%$, 0.12$\%$) &(0.23$\%$, 0.27$\%$)  &(0.085$\%$, 0.10$\%$)   &(0.19$\%$, 0.23$\%$) \\
& $\ >$ 0.005 &(0.11$\%$, 0.13$\%$) &(0.26$\%$, 0.29$\%$)  &(0.10$\%$, 0.12$\%$)   &(0.22$\%$, 0.26$\%$)    \\
&$\ >$  0.001 &(0.14$\%$, 0.17$\%$) &(0.35$\%$, 0.42$\%$)  &(0.14$\%$, 0.16$\%$)   &(0.33$\%$, 0.41$\%$)   \\
\hline
\end{tabular}
\caption*{\small{Same as Table 6, for the ellipticity error. The error in contaminated sources is caused by residual photon noise of LEOsat trails after trail removal ('repair').}}
\label{Table:NBstart}
\end{table}

The noise and systematic errors introduced by the bright trails can obscure the surface brightness fluctuations of galaxies, posing a significant challenge to the achievement of key scientific objectives. For example, in the photometric analysis of Type Ia supernovae (SNe Ia) and in exoplanet transit surveys, substantial contamination of flux or centroid measurements may render affected sources unusable, resulting in a permanent loss of valuable data. Similarly, errors in centroid and shape measurements can adversely impact weak gravitational lensing studies, particularly in the inference of cosmic shear.

Although the fraction of sources contaminated by satellite trails is relatively small (less than 0.50\%), it is imperative to identify and exclude these sources to ensure the accuracy of cosmic shear estimations and minimize systematic uncertainties. Moreover, the broad, low-brightness wings associated with satellite trails have not been fully accounted for in the current analysis. In reality, the width of these trails often extends well beyond the bright central region. The most significant contributor to the spatial extent of the PSF is atmospheric turbulence, which is typically modeled using von Kármán turbulence theory (\citealt{Walker2020, Hasan2022}). However, the extended structure of solar panels can also generate faint, diffuse wings in the trail profile.

Even after satellite trails are modeled and subtracted or masked, their residuals can contribute to the systematic error budget. Unlike stars and other masked sources that exhibit point symmetry, satellite trails possess line symmetry, introducing additional complexity. Cosmological measurements that rely on precise symmetry assumptions - particularly at low surface brightness levels - may be affected by these residuals \citep{Tyson2020}. Furthermore, when calculating shear in weak lensing through Fourier space, the impact of LEO satellite trails and their extended low-brightness wings must be carefully modeled and mitigated to preserve the integrity of shear measurements \citep{Zhang2015}.

\subsubsection{Result of slitless spectroscopic simulation}

The fraction of contaminated area in simulated slitless spectroscopic images is also found to be low. To evaluate the impact of LEOsat trails on slitless spectra, we analyzed a simulated GI-band slitless spectroscopic image with a size of 9000 pixels, containing 517 source samples. This assessment focused on both the contamination area and the level of noise introduced by LEOsat trails.

In slitless spectroscopy, both astronomical sources and satellite trails exhibit zeroth-order and first-order spectral components. Consequently, the widths (in pixels) of the zeroth-order and first-order components in the GI band were calculated under boundary thresholds of $0.1\,\mathrm{e}^{-}$, $0.5\,\mathrm{e}^{-}$, and $1.0\,\mathrm{e}^{-}$. 

\begin{itemize}  
\item For LEOsats at an orbital altitude of 550 km, the corresponding (zeroth-order, first-order) trail widths are (32.2, 341.3), (23.4, 322.5), and (21.1, 301.8), respectively.
\item For LEOsats at an altitude of 1200 km, the widths are (18.8, 335.5), (10.84, 309.2), and (9.1, 232.5), respectively. 
\end{itemize}

For each simulated spectroscopic image, we estimate the following levels of contamination under a population of $10^4$ LEOsats:

\begin{itemize}
    \item For 550 km altitude, the expected bright trail length across a single CCD is approximately 0.0037 degrees, corresponding to about 180 pixels. This yields a contamination rate of approximately 0.053\%, 0.049\%, and 0.046\% under the three threshold conditions, respectively.
    \item For 1200 km altitude, the expected trail length per CCD is about 0.010 degrees, leading to approximately 486.5 pixels, or a contamination rate of 0.136\%, 0.122\%, and 0.092\% under the three threshold conditions, respectively.
\end{itemize}

Assuming a total population of $10^5$ LEOsats distributed across altitudes ranging from 550 km to 1200 km, the contaminated area is estimated to be below 1.50\%.

\subsection{Mitigation method}

Several mitigation strategies have been proposed to minimize the impact of LEOsats on the operations of CSST. These strategies must be tailored to the specific characteristics of space-based telescopes, which differ significantly from those of ground-based observatories. For ground-based telescopes, a common mitigation approach involves temporarily closing the shutter during the passage of a satellite through the FoV. However, this method is not suitable for CSST, as the actuation of the shutter may introduce mechanical vibrations, potentially compromising the telescope’s pointing stability and accuracy.

One viable strategy for CSST involves optimizing the observation schedule to avoid periods or directions likely to be affected by LEOsat contamination. If a particular sky region is required for observation but is expected to be affected by flares, close satellite encounters, or a high density of LEOsats, the observation can be rescheduled or divided into shorter sub-exposures to reduce the risk of contamination. Accurate and up-to-date satellite ephemeris data are essential for the effective implementation of such strategies. In cases where transient astronomical events are of interest, observation plans must be adapted swiftly to accommodate the urgency of capturing such phenomena while still accounting for potential satellite interference. It is important to note that, unlike ground-based time-domain surveys, the CSST mission follows a predetermined and rigid scheduling framework, limiting the flexibility for major adjustments. However, observations may be shifted to adjacent time windows to minimize disruptions while preserving the integrity of the scientific objectives and overall mission efficiency.

An alternative mitigation strategy involves post-processing techniques, including trail subtraction, source marking, and adaptive weighting or masking. These approaches aim to identify and correct for the impact of satellite trails in the imaging data. In ground-based observations, modeling satellite trail profiles is complicated by atmospheric scintillation, which introduces irregularities along the trails. In contrast, space-based observatories like the CSST are not subject to such atmospheric effects, significantly simplifying trail modeling. For example, the Rubin Observatory must achieve a surface brightness precision on the order of one part in 10,000 for effective trail removal \citep{Walker2020}. Given that LEOsat trails in CSST images are substantially fainter than those in Rubin data, satisfying these precision requirements is more tractable for CSST. A profile model of the trail can be constructed to facilitate accurate subtraction. The distance between astronomical sources and satellite trails can also be used to evaluate potential contamination. Sources along or near trails can be flagged, and those exhibiting significant residual errors after trail subtraction can be masked. Figure 7 illustrates the relationship between the normalized distance of sources from the LEOsat trails (expressed as a multiple of the trail's FWHM) and the decimal logarithm of the relative increase in photometric error induced by either the trail signal or residual photon noise. The figure reveals a trend wherein photometric error diminishes with increasing distance from the trail. 
\begin{figure}
    \centering
	\includegraphics[width=0.48\linewidth]{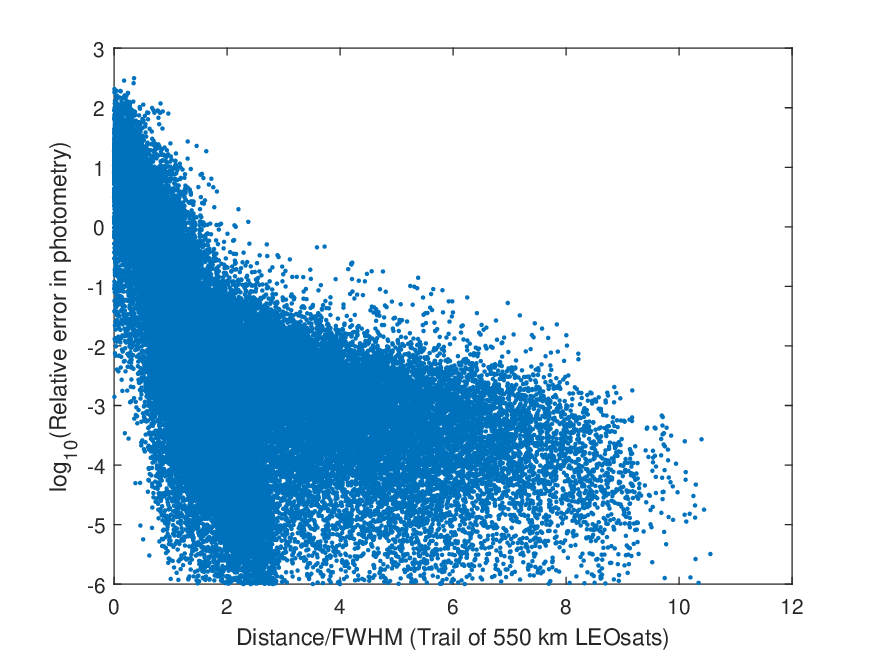}
    \hspace{0.01\linewidth}
	\includegraphics[width=0.48\linewidth]{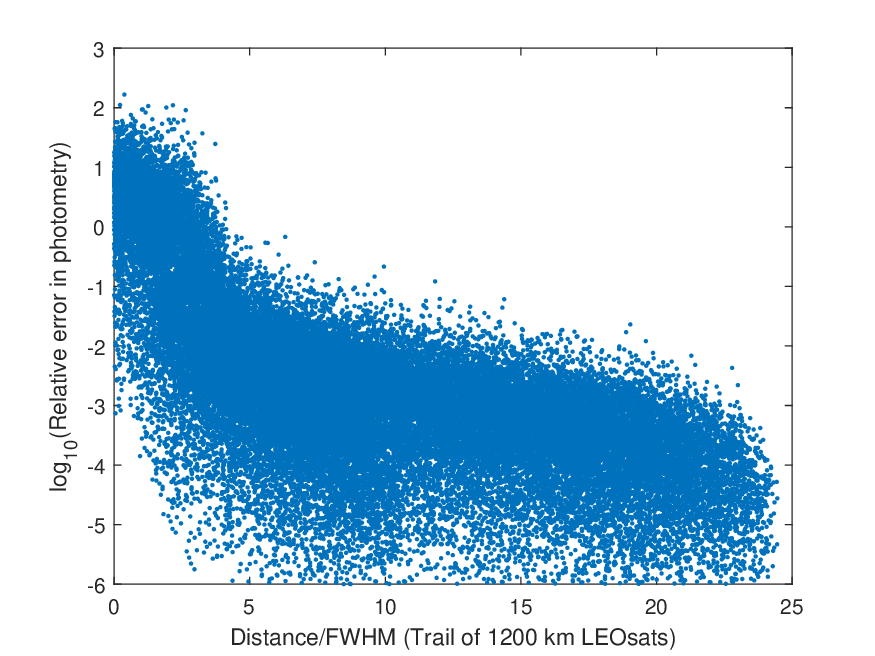}
	\includegraphics[width=0.48\linewidth]{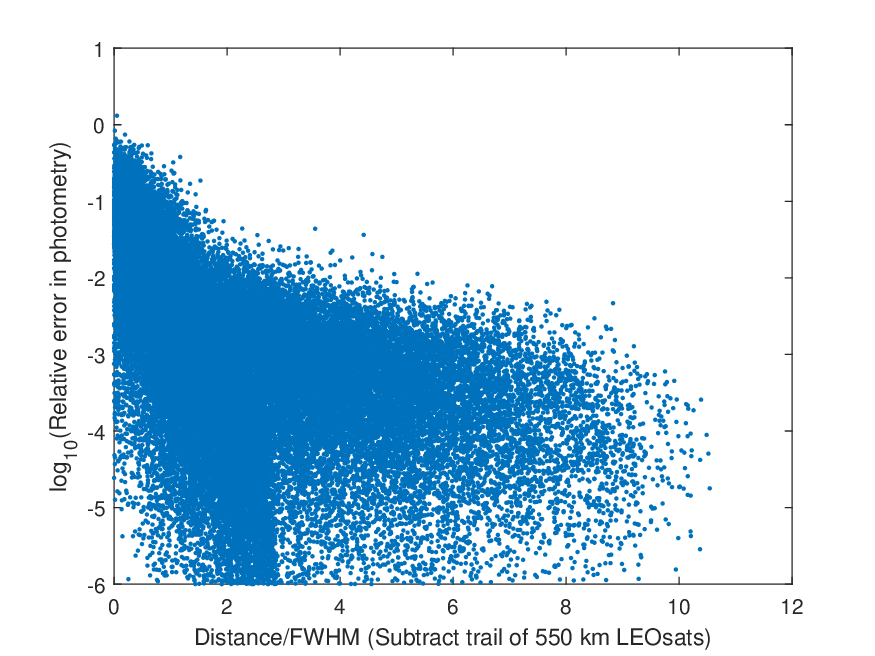}
	    \hspace{0.01\linewidth}
	\includegraphics[width=0.48\linewidth]{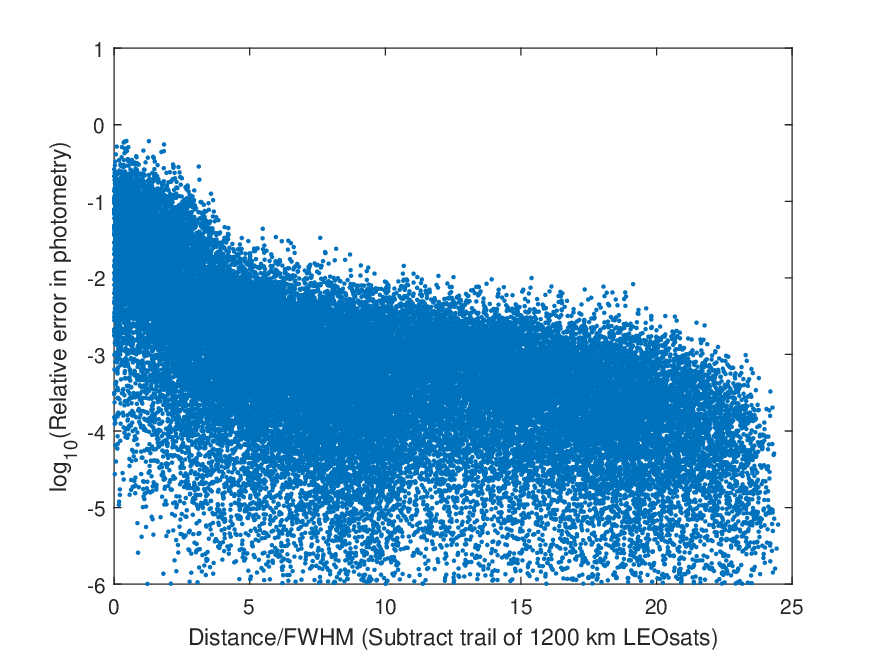}
 \caption{ 
The left panel demonstrates the relationship between the distance of sources from the LEOsat trails, normalized by the trail's FWHM, and the relative photometric error increase induced by the LEOsats at an altitude of 550 km. The right panel demonstrates this relationship for LEOsats at an altitude of 1200 km. The upper panel shows the residual trail, while the lower panel displays the result after trail subtraction and the remaining photon noise. We employ the decimal logarithm (log10) of the relative photometric error increase of contaminated sources to represent the distribution of the results. It is evident that the increase in photometric error decreases with increasing distance. }   
\end{figure}

\begin{figure}
\centering
	\includegraphics[width=0.48\linewidth]{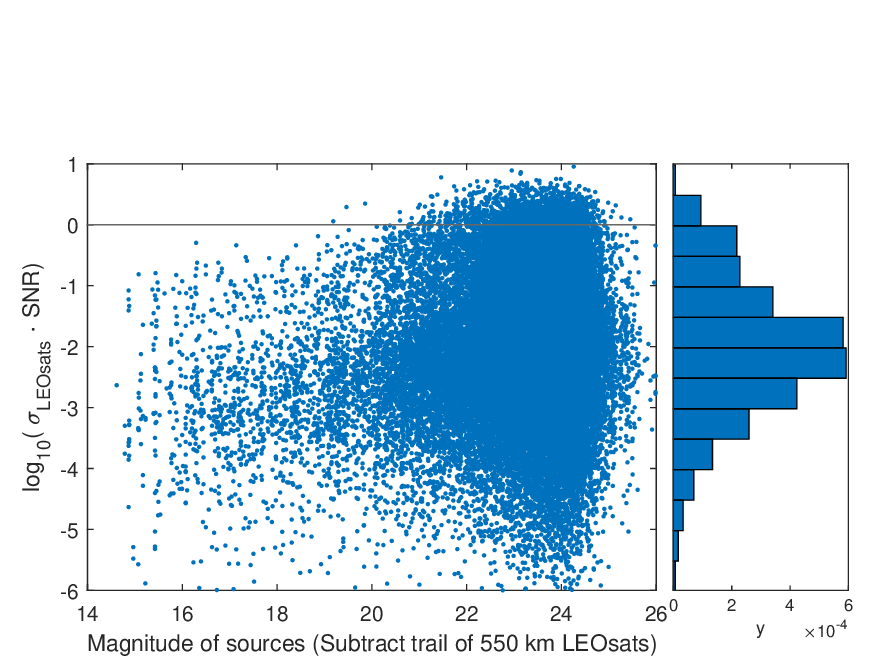}
    \hspace{0.01\linewidth}
	\includegraphics[width=0.48\linewidth]{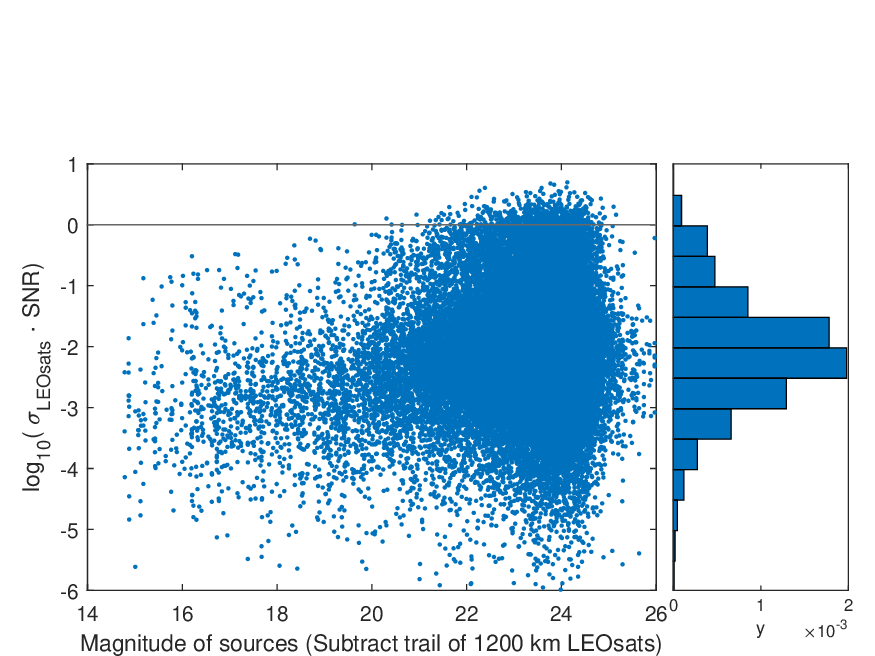}
 \caption{ A statistical comparison based on 2000 simulations is made between relative photometric errors caused by satellite trail photon noise and other noise sources, using sources with SNR $>$ 5. After trail correction, we compute the product of the trail-induced relative photometric error ($\sigma_{\rm LEOsats}$) and the source SNR, and take its decimal logarithm as the $y$-value. When compared with all sources in the image, the proportion of significantly affected sources remains very low. For $10^5$ LEOsats in 550 km orbits, during a 150-second exposure with an $i$-band CCD, 0.010\% of sources have $y>$ 0, and 0.055\% have $y>$ -1; for 1200 km, the corresponding values are 0.0095\%, and 0.096\%. These results indicate that, after correction, trail-induced photometric errors are negligible compared to other noise sources for bright sources. } 
\end{figure}

\section{Conclusion}

In this study, we investigate the impact of LEOsats on space-based astronomical observations, with a particular focus on the CSST. We compare the characteristics of LEOsat trail contamination in space-based versus ground-based telescopes, highlighting both the commonalities and key differences.

Using detailed simulations, we quantify the number, length, brightness, and profile of LEOsat trails within the CSST FoV. These trail characteristics are then incorporated into synthetic CSST images for further analysis. The simulations reveal that, for space-based observatories, LEOsat trails generally exhibit stable and moderate brightness, making them amenable to modeling and subtraction.

We assess the number of contaminated sources and quantify the increase in photometric, astrometric, and shape measurement errors induced by LEOsat trails. We further evaluate the effectiveness of trail subtraction—where residual photon noise remains but the trail signal is removed—in reducing contamination and mitigating measurement error. Even in scenarios involving up to $10^{5}$ LEOsats, the fraction of contaminated sources remains low, typically under 0.50\% in imaging data. For slitless spectroscopic images, the contaminated area is controlled to be below 1.50\%. After subtracting the LEOsat trails, residual photon noise causes relative photometric errors exceeding one-tenth of those from other error sources in approximately 0.10\% of all sources.

As shown in Figure 8, comparative analysis suggests that the measurement errors introduced by LEOsat trails, after subtraction ('repair'), are generally lower than those caused by background noise, as the trails predominantly affect the peripheries of sources. Furthermore, higher-altitude LEOsats (e.g., at 1200 km) are found to contaminate a greater number of sources than those at lower altitudes (e.g., at 550 km), both in absolute terms and as a proportion of total sources. Trail subtraction is shown to effectively reduce the number of contaminated sources and limit the accumulation of systematic errors. This initial work conducts a single-band simulation analysis using the HST F814W band, which is largely equivalent to the $i$-band in the CSST. The results in other bands are expected to follow similar patterns. Future work will extend these simulations to cover the entire range of CSST bands.

Several specific aspects warrant further investigation to enhance the mitigation of LEOsat-induced contamination in space-based astronomical data. In particular, data processing pipelines must be capable of accurately modeling satellite trails, restoring affected regions, and recovering transient astronomical sources compromised by such contamination. In our simulations, we adopt a satellite brightness threshold of V = 7 magnitude, consistent with recommendations from the astronomical community. For satellites exceeding this brightness, the trail flux is scaled proportionally to their increased brightness.

To mitigate the influence of LEOsat trails on source measurements, a weighting function can be applied to down-weight affected regions during data analysis. Additionally, proactive prediction and scheduling are essential for managing the impact of flares and close encounters. The profiles of satellite trails extend beyond the bright core into broad, low-surface-brightness wings, which require further characterization in the context of space-based imaging.

A comprehensive approach - encompassing improved modeling, predictive scheduling, and robust data processing - is necessary to minimize the impact of LEOsats on the scientific output of space telescopes. Future research should aim to address all relevant factors contributing to trail-induced contamination.

\begin{acknowledgements}
This work was supported by National Key R\&D Program of China No.2022YFF0503400. HT would like to thank Xiyang Fu, Weiyang Liu and Li Shao for their invaluable assistance and support throughout this work.
\end{acknowledgements}




\label{lastpage}

\end{document}